\documentclass{article}
\usepackage{spconf,amsmath,graphicx}

\usepackage{enumitem}
\usepackage{paralist}
\usepackage{multirow}
\setlist{nosep, leftmargin=14pt}
\usepackage{booktabs}
\usepackage[table]{xcolor}
\usepackage{array}

\usepackage{mwe} 
\usepackage{arydshln}

\usepackage[capitalize]{cleveref}
\usepackage{gensymb}
\crefname{section}{Sec.}{Secs.}
\Crefname{section}{Section}{Sections}
\Crefname{table}{Table}{Tables}
\crefname{table}{Tab.}{Tabs.}
\def\etal{\emph{et~al}. }

\title{Equi-ViT: Rotational Equivariant Vision Transformer for Robust Histopathology Analysis}
\name{
Fuyao Chen$^{1,2}$,
Yuexi Du$^{1}$,
El\'{e}onore V. Lieffrig$^{1}$,
Nicha~C.~Dvornek$^{1,3}$,
John~A.~Onofrey$^{1,3,4}$
}
\address{
    $^1$Departments of Biomedical Engineering, 
    $^2$Medical Scientist Training Program,\\
    $^3$Radiology \& Biomedical Imaging,
    $^4$Urology,
    Yale University, New Haven, CT, USA \\
}

\begin{document}
%
\maketitle
\begin{abstract}
Vision Transformers (ViTs) have gained rapid adoption in computational pathology for their ability to model long-range dependencies through self-attention, addressing the limitations of convolutional neural networks that excel at local pattern capture but struggle with global contextual reasoning. Recent pathology-specific foundation models have further advanced performance by leveraging large-scale pretraining. However, standard ViTs remain inherently non-equivariant to transformations such as rotations and reflections, which are ubiquitous variations in histopathology imaging. To address this limitation, we propose Equi-ViT, which integrates an equivariant convolution kernel into the patch embedding stage of a ViT architecture, imparting built-in rotational equivariance to learned representations. Equi-ViT achieves superior rotation-consistent patch embeddings and stable classification performance across image orientations. Our results on a public colorectal cancer dataset demonstrate that incorporating equivariant patch embedding enhances data efficiency and robustness, suggesting that equivariant transformers could potentially serve as more generalizable backbones for the application of ViT in histopathology such as digital pathology foundation models.
\end{abstract}
\begin{keywords}
Digital Histopathology, Vision Transformer, Rotation Equivariance, Artificial Intelligence

\end{keywords}
\section{Introduction}
\label{sec:intro}

Histopathology images present significant challenges for deep learning models due to the arbitrary orientations of tissue structures, making rotation-invariance a highly desirable property for robust analysis. Conventional Convolutional Neural Networks (CNNs) typically rely on geometric data augmentation to handle such variability, but this does not guarantee true rotation-robust performance and often requires test-time augmentation~\cite{Lafarge2020_SE(2),Veeling2018_RECNN}.
Recent work to address these challenges proposed using an equivariant convolution kernel design, Gaussian Mixture Ring Convolution (GMR-Conv)~\cite{Du2025_GMR}.
GMR-Conv uses a set of Gaussian-weighted rings to achieve rotation and reflection equivariance without incurring extra computational cost. GMR-Conv mitigates the discretization errors of prior circular kernels, preserving symmetry more faithfully, and it was shown to match or surpass conventional CNN performance on orientation-agnostic image tasks while being more robust and efficient than previous state-of-the-art equivariant methods~\cite{Weiler2019_e2CNN}.

Meanwhile, Vision Transformers (ViTs)~\cite{Dosovitskiy2020_ViT} have surged in popularity for computational pathology due to their ability to model long-range dependencies via self-attention, addressing the limitation of CNNs, which excel at local pattern capture but struggle with global context~\cite{Nechaev2024_vitpath}, especially for high-resolution pathology images. Large ViT-based foundation models pretrained on massive histopathology datasets have recently delivered state-of-the-art results, outperforming models without pathology-specific pretraining on classification and segmentation benchmarks~\cite{Breen2024_vit_foundation}. 
Despite their success, standard ViTs lack inherent equivariance to rotation and reflection transformations, which are ubiquitous in histopathology slides, motivating recent efforts to incorporate group-equivariant mechanisms into ViTs~\cite{Romero2020_GSNet,Xu2023_E2ViT}.

Existing equivariant ViTs address symmetry through either positional encoding or attention. For example, Romero \etal~\cite{Romero2020_GSNet} generalize the self-attention mechanism by lifting features from the original spatial domain to a joint domain that includes group elements, so that it becomes equivariant to a symmetry group. 
Furthermore, Xu \etal~\cite{Xu2023_E2ViT} proposed an $E(2)$-Equivariant ViT (GE-ViT) that approaches equivariance via a group-equivariant positional encoding.
However, the equivariance of image tokenization is neglected in these works.
Although group-equivariant self-attention substitutes ViT’s patch embedding with a lifting self-attention embedding, the initial linear projections remain unconstrained and may disrupt equivariant representations.

In this work, we address this limitation by introducing an equivariant patch embedding. We present an Equivariant Vision Transformer (Equi-ViT) for histopathology that integrates GMR-Conv into the patch embedding stage, endowing the ViT architecture with built-in rotational and reflection equivariance for input features from the beginning. Our approach aims to improve efficiency and robustness in analyzing pathology images. Equi-ViT can serve as a robust backbone for future foundation models
in digital pathology, providing orientation-invariant feature representations.

\section{Methods}

\paragraph*{Equivariant Feature Learning.} 
Gaussian Mixture Ring convolution (GMR-Conv)~\cite{Du2025_GMR} is a kernel design that achieves rotation and reflection equivariance through radial symmetry. Instead of a conventional square filter with independent kernel elements, GMR-Conv is constructed as a weighted combination of concentric Gaussian rings, which smooths the radial profile of the kernel. This kernel design mitigates the discretization errors that normally break equivariance in standard non-smooth radially symmetric filters, thereby preserving robust rotation and reflection equivariance in the convolution’s output features. Furthermore, GMR-Conv implements equivariance in a computational and parameter-efficient manner without any substantial additional computation overhead.

\paragraph*{Equivariant Vision Transformer (Equi-ViT).} 
In our Equi-ViT model, we integrate the GMR-Conv kernel by replacing the standard convolution in the patch embedding stage with a GMR-based convolution layer. The patch embedding layer of ViT ordinarily projects image patches into token embeddings using a linear projection (implemented as a single convolution layer with kernel size equal to the patch size), which is not equivariant.
In contrast, we intend to extract equivariant patch features using GMR-Conv layers for feature embedding. The plug-and-play nature of GMR-Conv allows it to be inserted seamlessly without altering the overall ViT architecture. Specifically, we adopted the ViT-Base backbone implemented in the Hugging Face Transformers library as our base architecture. As a result, ViT’s initial patch tokens carry orientation-equivariant image features, making the model’s representation robust to arbitrary rotations of input.

\section{Experiments and Results}
\label{sec:experiments and results}

\subsection{Experimental Setup}
\paragraph*{Data.}
Data from the \textbf{NCT-CRC-HE-100K}~\cite{Kather2019_nctcrc} public dataset comprises 100{,}000 hematoxylin-and-eosin (H\&E) stained image patches of colorectal cancer and normal colon tissue, curated from formalin-fixed paraffin-embedded whole-slide images
~\cite{Kather2019_nctcrc}. 
Each non-overlapping patch is $224\times224$ pixels, extracted at $0.5\,\mu\text{m/pixel}$, and annotated by expert pathologists into nine tissue categories: adipose tissue (ADI), background (BACK), debris (DEB), lymphocytes (LYM), mucus (MUC), smooth muscle (MUS), normal mucosa (NORM), cancer-associated stroma (STR), and colorectal adenocarcinoma epithelium (TUM). The dataset includes a predefined validation subset of 7{,}180 images provided by the data source to ensure consistent benchmarking across studies. Experiments used the provided patch-level labels without additional relabeling and performed stratified splits by tissue class for training and validation.

\paragraph*{Baseline Models.}
We compared Equi-ViT to both non-equivariant and equivariant ViT models as well as traditional CNNs.
We use conventional \textbf{ViT} as the baseline model. Additionally, we augment the ViT model with standard Conv2D kernels of sizes 6 and 11 (\textbf{Conv ViT}) for patch embedding.
We compare to the state-of-the-art equivariant learning methods for patch embedding, using the R2Conv kernel from E(2)-WRN16~\cite{Weiler2019_e2CNN} (\textbf{E(2) ViT}), which constrains its filters to an equivariant kernel subspace.
Finally, we compared our results to various CNN-based structures and their group equivariant variations, including conventional ResNet18 (\textbf{R18})~\cite{He2015_R18}, the Wide ResNet16 (\textbf{E(2)-WRN16})~\cite{Weiler2019_e2CNN}, and the GMR-Conv enabled ResNet18 (\textbf{GMR-R18})~\cite{Du2025_GMR}.

\paragraph*{Implementation.}
Our Equi-ViT design employs two sequential GMR-Conv kernels of sizes 6 and 11 to replace the original patch embedding mechanism in the standard ViT architecture. 
We trained all models using the AdamW optimizer and a cosine annealing learning rate scheduler, with an initial learning rate of \(5\times10^{-5}\). Training was performed for 10 epochs with a batch size of 64, and cross-entropy loss was used for the multi-class classification task. All implementations were based on the PyTorch framework and executed on an NVIDIA A5000 GPU. Consistent with standard practice in equivariant feature learning studies~\cite{Cohen2016_GECNN,Dudar2018_symCNN,Mo2022_ricCNN,Weiler2019_e2CNN}, no geometric data augmentation was applied during training to isolate and demonstrate the intrinsic capabilities of equivariant representations without confounding effects.

\paragraph*{Evaluation Metrics.} 
We evaluate model performance for each dataset by computing the classification accuracy on: (1) the original test set without rotation; and (2) a rotated test set, where test images are rotated in 10{\textdegree} increments. The rotation accuracy is reported as mean$\pm$standard deviation across all rotations. We assess significant differences (\(\alpha = 0.05\)) between models using two-tailed paired \(t\)-tests.

\subsection{Results}

\begin{figure}[t]
    \centering
    \includegraphics[width=1.0\columnwidth]{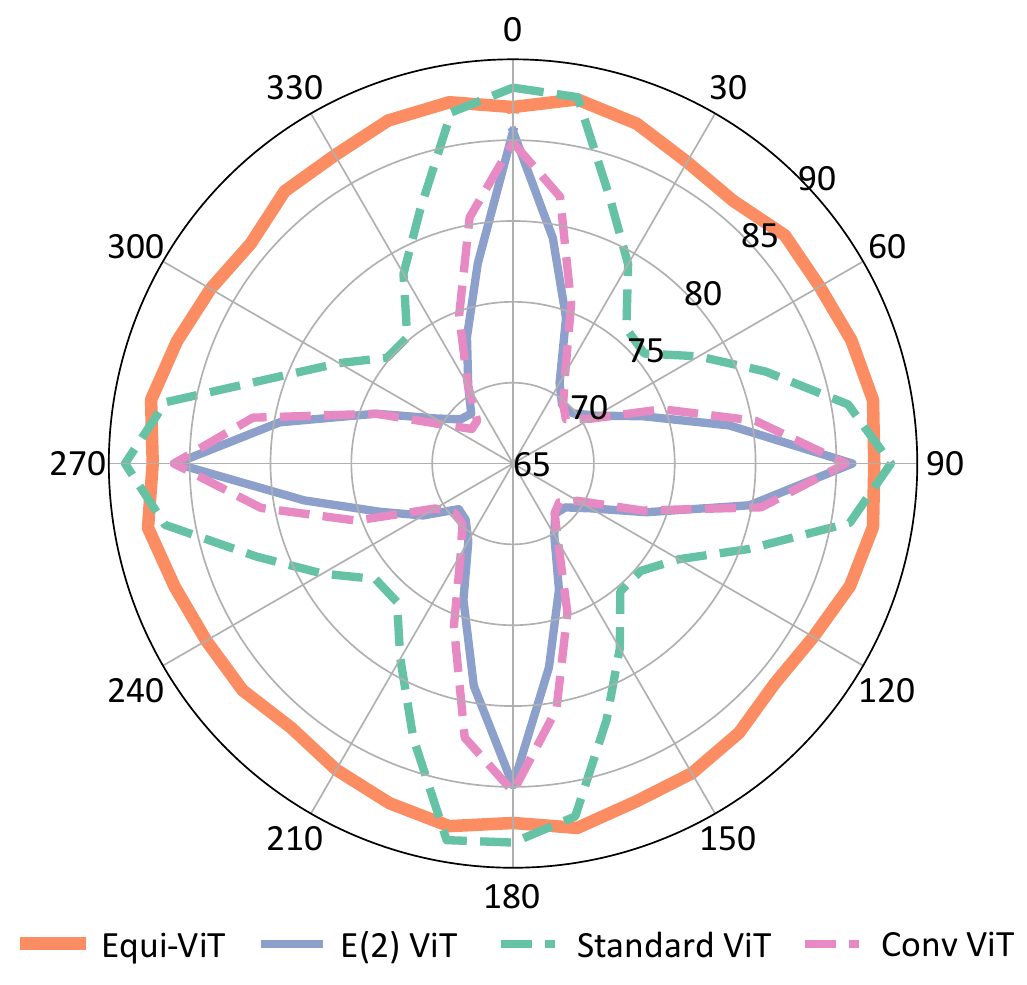}
    \vspace{-2.0\baselineskip}
    \caption{\textbf{Classification Accuracy Across Rotation Angles}. We visualize the classification accuracy (\%) per test rotation angles for both equivariant methods Equi-ViT (our approach) and ViT with R2Conv kernels~\cite{Weiler2019_e2CNN} (E(2) ViT) and for non-equivariant methods Standard-ViT and ViT with standard Conv2D kernels (Conv ViT). We scale the radial axis to better visualize differences in performance.}
    \label{fig:radar}
    \vspace{-1.0\baselineskip}
\end{figure}

\paragraph*{Rotation Evaluation.}
Equi-ViT demonstrates superior rotation equivariance in histopathology classification tasks (Rot. Acc. 86.8$\pm$0.59)  when compared to both the standard ViT (83.1$\pm$6.93) and the Conv2D ViT model (77.6$\pm$7.32) (\cref{fig:radar} and \cref{tab:rotation_comparison}). 
Furthermore, Equi-ViT outperforms the alternative equivariant embedding approach, E(2) ViT (74.5$\pm$5.1).
This is likely due to E(2) ViT's equivariance being limited to discrete cyclic rotation subgroups, such as $C_4$ (i.e., 90{\textdegree} increments), making it challenging to generalize across intermediate rotations not aligned with the predefined group. 
Equi-ViT exhibits consistent classification performance across various rotation angles, showcasing its robustness and reliability in handling rotational variations in the input pathology images.

Compared to CNN-based approaches, our proposed Equi-ViT model does not surpass the performance of CNNs, particularly GMR-R18.
This discrepancy is likely attributable to the higher demand for training data necessitated by the heavily parameterized ViT architecture. However, the Equi-ViT model demonstrated more stable classification performance across rotations as compared to R18 and E(2)-WRN16. Furthermore, our Equi-ViT design, incorporating GMR-Conv, demonstrates efficient equivariant patch token embedding, with the embedding module consisting of 0.79M parameters and occupying 3.0 MB of memory. This is in contrast to Conv ViT, which employs Conv2D kernels in a comparable embedding module structure but comprises 2.4M parameters and requires 9.1 MB of memory, thereby underscoring the parameter efficiency of Equi-ViT's design.

\begin{table}[t]
\centering
\caption{\textbf{Evaluation Results.} Classification accuracy reported for test set images in their Original orientation and the Rotated test set (mean$\pm$SD over all angles).
We highlight the best performance with bold. Our model is shaded in gray.}
\setlength{\tabcolsep}{6pt}
\resizebox{\columnwidth}{!}
{
    \begin{tabular}{llcccc}
    \toprule
    \textbf{Arch.} & \textbf{Model} & \textbf{\#Param.} & \textbf{Memo.} & \textbf{Orig. (\%)} & \textbf{Rot. (\%)} \\ 
    \midrule
    \multirow{3}{*}{CNN} & R18 & 11.2M & \textbf{3.4G} & 93.7 & 87.3 $\pm$ 5.1\\
    & E(2)-WRN16 & 10.8M & 20.9G & 93.8 & 92.5 $\pm$ 3.5\\
    & GMR-R18 & \textbf{3.9M} & 6.2G & \textbf{95.6} & \textbf{95.2 $\pm$ 0.2} \\
    \midrule
    \multirow{4}{*}{ViT} & ViT & \textbf{85M} & \textbf{10.8G} & \textbf{88.2} & 83.1 $\pm$ 6.9\\ 
    & Conv ViT & 87M & 11.0G & 84.8 & 77.6 $\pm$ 7.3\\ 
    & E(2) ViT & 94M & 28.4G & 85.5 & 74.5 $\pm$ 5.1 \\
    \cmidrule{2-6}
    & \cellcolor[HTML]{EFEFEF}Equi-ViT & \cellcolor[HTML]{EFEFEF}86M & \cellcolor[HTML]{EFEFEF}10.9G & \cellcolor[HTML]{EFEFEF}87.0 & \cellcolor[HTML]{EFEFEF}\textbf{86.8 $\pm$ 0.6}\\ 
    \bottomrule
    \end{tabular}
}
\label{tab:rotation_comparison}
\vspace{-1.2\baselineskip}
\end{table}

\begin{figure*}[t]
    \centering
    \includegraphics[width=0.95\textwidth]{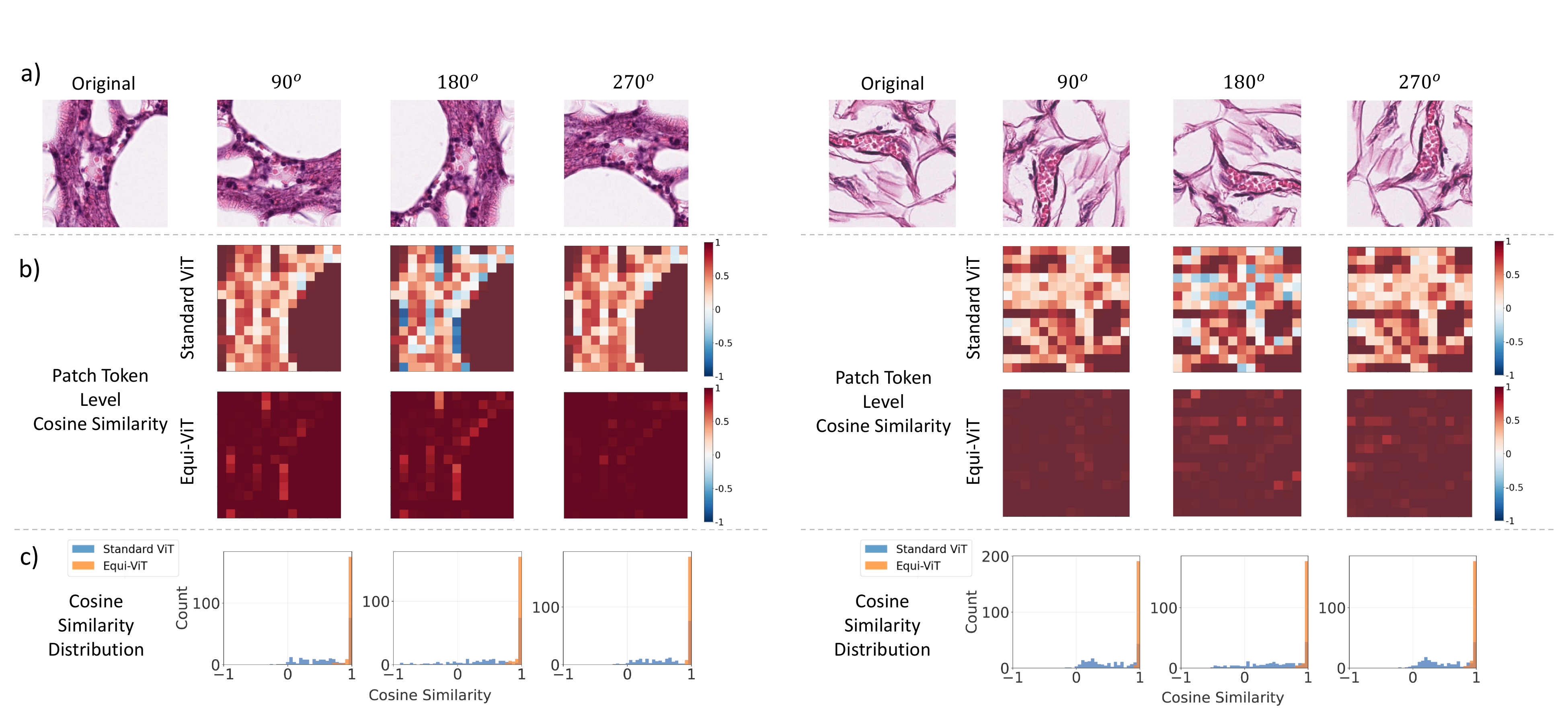}
    \vspace{-1.5\baselineskip}
    \caption{\textbf{Patch Token Embedding Equivariance Analysis} (a) Two example histopathology images undergoing 90{\textdegree}, 180{\textdegree}, and 270{\textdegree} rotation. (b) Cosine similarity was calculated between the corresponding patch token features extracted from original angle image and the rotated images, compared between standard ViT and Equi-ViT. (c) The distribution of cosine similarity values of patch tokens for standard ViT and Equi-ViT demonstrates that our equivariant feature embedding approach maintains nearly perfect patch equivariance (cosine values equal to 1) across rotations.}
    \label{fig:cos_sim}
    \vspace{-1.0\baselineskip}
\end{figure*}

\paragraph*{Patch Token Embedding Equivariance Analysis.}
To demonstrate the robustness of equivariant feature extraction with Equi-ViT, we extracted patch tokens from images rotated by 90, 180, and 270 degrees. We then rotated the tokens and patch grid back to their original orientation and calculated the cosine similarity between the patch token features extracted from the rotated and original orientation images. Our results~(\cref{fig:cos_sim}) indicate that patch token features extracted using Equi-ViT exhibited superior alignment between rotated and original orientation images. In contrast, patch-level token features extracted with the standard ViT were less aligned, often orthogonal, and in some cases, inversely correlated between the rotated and original images, indicating the structural features might be corrupted, and may further impact the following transformer blocks.

\paragraph*{Ablation Studies.}
We compared various kernel types and size combinations for the patch embedding process. As the kernel size of the GMR-Conv increases, its effective receptive field expands, and the fine-grained rotational sensitivity becomes coarser due to its smooth radial symmetry. This characteristic may explain the suboptimal performance observed when directly replacing the patch embedding kernel with a $16\times16$ GMR-Conv kernel. Subsequent analyses~(\cref{tab:vit_gmr_variants}) using smaller kernel combinations consistently demonstrated robust pathology classification performance across different image rotations, as evidenced by the low accuracy standard deviation across various rotation angles. The highest performing model utilized sequential GMR-Conv kernel sizes of 6 and 11 (denoted [6, 11]) for patch token embedding, achieving a rotational accuracy of 86.8$\pm$0.59. When these GMR-Conv kernels were replaced with Conv2D kernels of identical sizes (Conv [6, 11]), a significant decrease in rotational accuracy was observed, dropping to 77.6$\pm$7.32.

\begin{table}[t]
\vspace{-0.\baselineskip}
\caption{\textbf{Ablation Study Results.} Classification accuracy reported for test set images in their Original orientation and the Rotated test set (mean$\pm$SD accuracy over all angles) using different kernel configurations.
We highlight the best performance with bold. Our model is shaded in gray.}
\centering
\setlength{\tabcolsep}{6pt}
\resizebox{\columnwidth}{!}
{
    \begin{tabular}{lccc}
    \toprule
    \textbf{Embedding} & \textbf{Kernel Config.} & \textbf{Orig. (\%)} & \textbf{Rot. (\%)} \\ 
    \midrule
    Conv. & [6, 11] & 84.8 & 77.6 $\pm$ 7.32 \\ 
    \midrule
    \multirow{6}{*}{GMR} 
    & \cellcolor[HTML]{EFEFEF}[6, 11] & \cellcolor[HTML]{EFEFEF}\textbf{87.0} & \cellcolor[HTML]{EFEFEF}\textbf{86.8 $\pm$ 0.59} \\ 
    & [8, 9] & 85.7 & 86.5 $\pm$ 0.47 \\
    & [10, 7] & 84.0 & 83.1 $\pm$ 1.11 \\
    & [12, 5] & 86.2 & 85.4 $\pm$ 1.31 \\
    & [6, 6, 6] & 81.3 & 81.0 $\pm$ 1.16 \\
    & [16] & 84.4 & 74.3 $\pm$ 7.40 \\
    \bottomrule
    \end{tabular}
}
\label{tab:vit_gmr_variants}
\vspace{-1\baselineskip}
\end{table}

\section{Discussion and Conclusion}
\label{sec:discussion}

In this study, we introduce Equi-ViT, which integrates Gaussian Mixture Ring (GMR) Convolution into the patch token embedding stage of the Vision Transformer (ViT) to enhance rotation equivariance. This design is ideal for histopathology analysis, where orientation variability in tissue morphology are ubiquitous. Equi-ViT provides reliable, rotation-invariant token representations, essential for accurate classification. The model achieved a rotation accuracy of 86.8$\pm$0.59, demonstrating the effectiveness of equivariant patch embedding in enhancing rotation-consistent representation learning and its importance for developing geometrically robust transformers.

Prior frameworks such as GSNet~\cite{Romero2020_GSNet} and GE-ViT~\cite{Xu2023_E2ViT} achieve mathematically guaranteed equivariance by redefining self-attention via lifting tokens into a joint space of spatial positions and orientations and introducing group-equivariant positional encoding. However, these guarantees are typically limited to discrete or predefined rotation groups, restricting scalability and flexibility. Furthermore, while GE-ViT, built on the basis of GSNet architecture, introduces a conceptually and theoretically grounded approach to group-equivariant attention, its current implementation poses computational challenges. The necessity of local lifting and high-dimensional tensor operations for practical model realization can lead to significant memory requirements and inefficient GPU utilization. 
These effects make training intractable for higher-resolution and larger-size histopathological images, such as the NCT-CRC data, due to the model's 3D convolution operations and irregular tensor shape increasing the intermediate tensor dimension, requiring 89.6GB of initial memory allocation. 
Furthermore, the accuracy performance for patch embedding using the R2Conv group-equivariant kernel~\cite{Weiler2019_e2CNN} was likely impacted by its restriction to discrete rotation subgroups, which makes it challenging to approximate features at intermediate orientations. Expanding the subgroup size can potentially improve angular approximation, but at the cost of increased memory and parameter demands, limiting overall computational efficiency (\cref{tab:rotation_comparison}).

While our Equi-ViT design demonstrated superior performance compared to the baseline ViT models, it did not surpass the performance of CNN-based equivariant models (\cref{tab:rotation_comparison}). This can be attributed to the need for larger training datasets, owing to ViT's high parameterization and weak inductive biases. Unlike CNNs, ViTs must learn spatial structures and invariances directly from data. Nevertheless, Equi-ViT has the potential to address these challenges by providing stronger spatial priors, which can enhance training efficiency and data utilization. 
While the proposed GMR-based embedding introduces feature rotation equivariance efficiently, further work could focus on integration with radial or relative positional embeddings, rotation-consistent attention biases, and group-aware regularization strategies to enhance our current work and approximate full equivariance in a more computationally efficient and scalable manner for medical image analysis applications.
Future work is also warranted to evaluate the model on a broader range of histopathology datasets to improve its potential to serve as a foundation model for robust and accurate histopathology analysis.


\section{Compliance with Ethical Standards}
This study was conducted retrospectively using human subject data made available in open access by Kather et al. \cite{Kather2019_nctcrc}. Ethical approval was not required, as confirmed by the license attached to the open-access data.

\section{Acknowledgments}
\label{sec:acknowledgments}

F.C. was supported by NIH Medical Scientist Training Program Training Grant T32GM007205.
The authors have no relevant financial or non-financial interests to disclose.

\bibliographystyle{IEEEbib}
\footnotesize
\bibliography{refs}

\end{document}